\documentclass[article,aps,reprint,amsmath,amssymb,superscriptaddress]{revtex4-2}

\usepackage[a4paper, left=50pt, right=35pt, top=65pt, bottom=65pt, headsep=2pt,marginparwidth=66pt]{geometry}
\usepackage{graphicx}
\usepackage{overpic}
\usepackage{float}
\usepackage{appendix}
\usepackage{array}
\usepackage{subfigure}
\usepackage{epsfig,graphicx,epstopdf}
\usepackage[colorlinks,
            linkcolor=blue,
            urlcolor=blue,
            filecolor=blue,
            anchorcolor=blue,
            citecolor=blue, ]{hyperref}

\usepackage{enumitem}
\usepackage{makecell}
\usepackage{color}
\usepackage{lineno}
\usepackage{booktabs}

\usepackage[english]{babel}

\begin{document}

\title{Measurement of attenuation length of the muon content in extensive air showers from 0.3 to 30 PeV with LHAASO}

\date{\today}

\author{Zhen Cao}
\affiliation{Key Laboratory of Particle Astrophysics $\&$ Experimental Physics Division $\&$ Computing Center, Institute of High Energy Physics, Chinese Academy of Sciences, 100049 Beijing, China}
\affiliation{University of Chinese Academy of Sciences, 100049 Beijing, China}
\affiliation{TIANFU Cosmic Ray Research Center, Chengdu, Sichuan,  China}
 
\author{F. Aharonian}
\affiliation{Dublin Institute for Advanced Studies, 31 Fitzwilliam Place, 2 Dublin, Ireland }
\affiliation{Max-Planck-Institut for Nuclear Physics, P.O. Box 103980, 69029  Heidelberg, Germany}
 
\author{Axikegu}
\affiliation{School of Physical Science and Technology $\&$  School of Information Science and Technology, Southwest Jiaotong University, 610031 Chengdu, Sichuan, China}
 
\author{Y.X. Bai}
\affiliation{Key Laboratory of Particle Astrophysics $\&$ Experimental Physics Division $\&$ Computing Center, Institute of High Energy Physics, Chinese Academy of Sciences, 100049 Beijing, China}
\affiliation{TIANFU Cosmic Ray Research Center, Chengdu, Sichuan,  China}
 
\author{Y.W. Bao}
\affiliation{School of Astronomy and Space Science, Nanjing University, 210023 Nanjing, Jiangsu, China}
 
\author{D. Bastieri}
\affiliation{Center for Astrophysics, Guangzhou University, 510006 Guangzhou, Guangdong, China}
 
\author{X.J. Bi}
\affiliation{Key Laboratory of Particle Astrophysics $\&$ Experimental Physics Division $\&$ Computing Center, Institute of High Energy Physics, Chinese Academy of Sciences, 100049 Beijing, China}
\affiliation{University of Chinese Academy of Sciences, 100049 Beijing, China}
\affiliation{TIANFU Cosmic Ray Research Center, Chengdu, Sichuan,  China}
 
\author{Y.J. Bi}
\affiliation{Key Laboratory of Particle Astrophysics $\&$ Experimental Physics Division $\&$ Computing Center, Institute of High Energy Physics, Chinese Academy of Sciences, 100049 Beijing, China}
\affiliation{TIANFU Cosmic Ray Research Center, Chengdu, Sichuan,  China}
 
\author{W. Bian}
\affiliation{Tsung-Dao Lee Institute $\&$ School of Physics and Astronomy, Shanghai Jiao Tong University, 200240 Shanghai, China}
 
\author{A.V. Bukevich}
\affiliation{Institute for Nuclear Research of Russian Academy of Sciences, 117312 Moscow, Russia}
 
% ...
\collaboration{The LHAASO Collaboration}

\email{fengxt@mail.sdu.edu.cn \\ fengcf@sdu.edu.cn \\ hyzhang@ynu.edu.cn\\llma@ihep.ac.cn}

\begin{abstract}
The attenuation length of the muon content in extensive air showers provides important information regarding the generation and development of air showers. This information can be used not only to improve the description of such showers but also to test fundamental models of hadronic interactions. Using data from the LHAASO-KM2A experiment, the development of the muon content in high-energy air showers was studied. The attenuation length of muon content in the air showers was measured from experimental data in the energy range from 0.3 to 30 PeV using the constant intensity cut method. By comparing the attenuation length of the muon content with predictions from high-energy hadronic interaction models (QGSJET-II-04, SIBYLL 2.3d, and EPOS-LHC), it is evident that LHAASO results are significantly shorter than those predicted by the first two models (QGSJET-II-04 and SIBYLL 2.3d) but relatively close to those predicted by the third model (EPOS-LHC). Thus, the LHAASO data favor the EPOS-LHC model over the other two models. The three interaction models confirmed an increasing trend in the attenuation length as the cosmic-ray energy increases.

\end{abstract}

\maketitle

\section{Introduction}

In recent years, considerable attention has been paid to the disparities between theoretical models for extensive air shower (EAS) development implemented in Monte Carlo (MC) simulations and observations made in EAS experiments. These differences are crucial indicators for improving theoretical models.

In cosmic-ray air showers, muons are generated through the decay of $\pi$ and $K$ mesons ($\pi^{\pm}$ $\rightarrow$ $\mu^{\pm}$ + ${\nu}_{\mu}$($\bar{\nu}_{\mu}$), $K^{\pm}$ $\rightarrow$ $\mu^{\pm}$ + $\nu_{\mu}$($\bar{\nu}_{\mu}$)) generated in hadronic collisions in the early stages of an EAS. Muons possess high penetration capabilities and preserve crucial information about hadronic interactions at very high energies. Therefore, the muon content of an air shower and its attenuation length are sensitive indicators for validating hadronic interaction models.

The Pierre Auger Observatory has observed an excess of muon content in air showers with primary energy ranging from 6 to 16 EeV, which is higher than that predicted using LHC-tuned hadronic models~\cite{PAOtest}. This excess muon content within the same energy range was also confirmed by the Telescope Array (TA) experiment, which is the largest experiment conducted in the Northern Hemisphere~\cite{PhysRevD.98.022002}. Additionally, measurements conducted by the Sydney University Giant Air-shower Recorder (SUGAR), using a muon detector (MD), have shown an increasing excess of muons starting from 0.1 EeV and continuing with increasing primary energy~\cite{SUGAR}. In contrast, IceTop reported smaller muon densities than the post-LHC model predictions for energies ranging from 1 to 100 PeV~\cite{icetopden}. The KASCADE-Grande observatory has also measured the attenuation length of the muon content of air showers at sea level for energies between 20 and 100 PeV and found that it is underestimated by all high-energy hadronic interaction models currently available~\cite{KASCADEatt}. Therefore, it can be concluded that none of the post-LHC high-energy hadronic interaction models can consistently describe the evolution of muon content in EASs in the aforementioned energy regime.

Understanding the disparities in muon content of EASs is crucial for understanding the discrepancies between model predictions and observations. The Large High Altitude Air Shower Observatory (LHAASO), equipped with 1188 soil-buried MDs, provides unparalleled precision in measuring muon content. This study specifically focused on determining the attenuation length of muon content in very-high-energy EASs ranging from 0.3 to 30 PeV, which encompasses the knee region of the all-particle energy spectrum.

The remainder of this paper is organized as follows. Section~\ref{sec:TWO} describes the experimental setup and event information for the LHAASO array, including the performance of MDs, principles of muon content measurement, generation of simulations, and event selection criteria. In Section~\ref{sec:Three}, the attenuation lengths of muon content were assessed using the constant intensity cut (CIC) method. Additionally, the attenuation lengths of simulated data (EPOS-LHC, QGSJET-II-04, and SIBYLL 2.3d) were measured, and the variation trends in the attenuation length with the integral flux and energy are presented. We also provide a detailed analysis of the sources of systematic errors. Finally, Section~\ref{sec:FOUR} presents the conclusions.

\section{Experiments and simulations\label{sec:TWO}}
\subsection{LHAASO KM2A detector}

\begin{figure}[htpb]
\centering
\includegraphics[width=7cm]{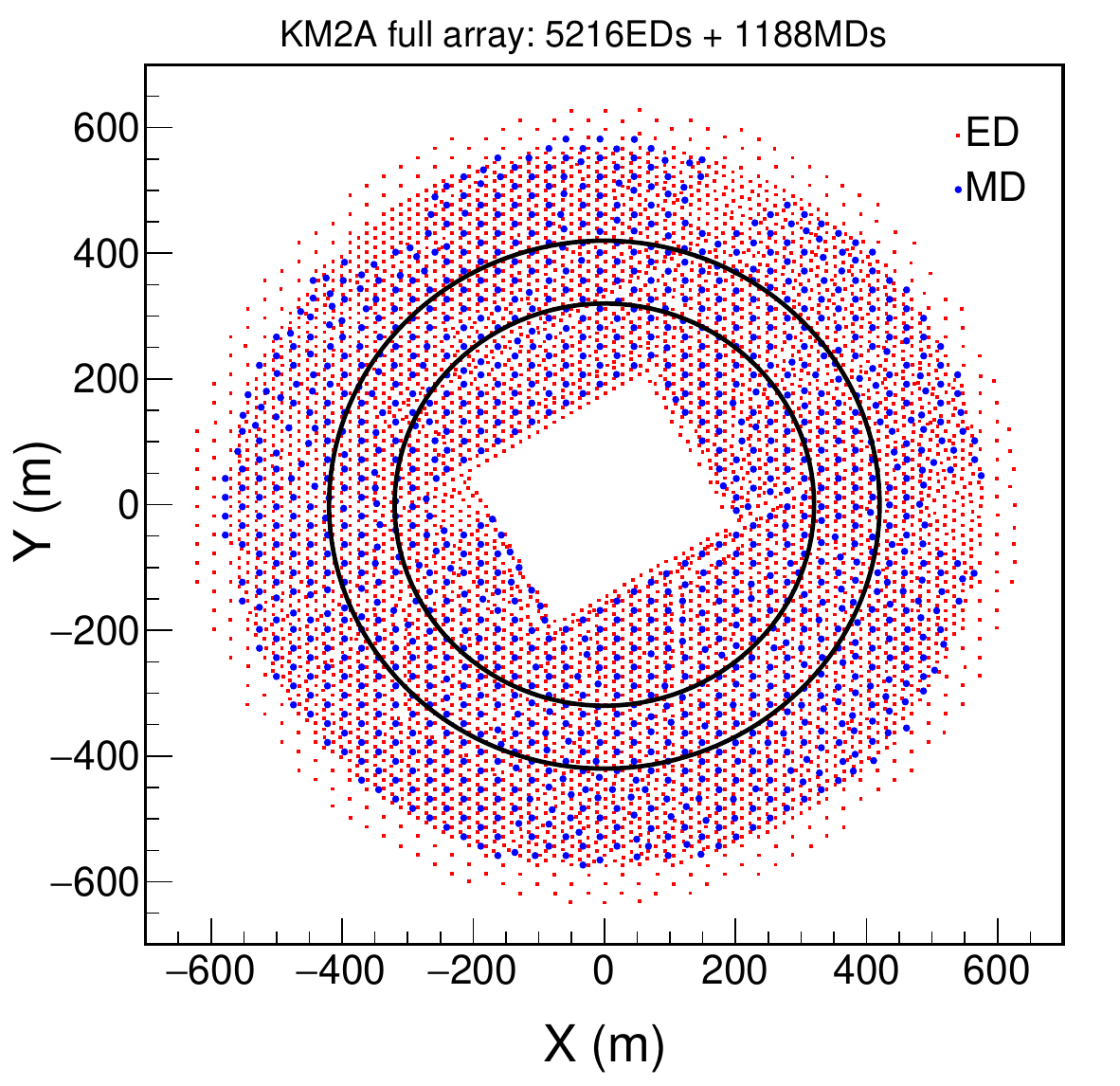}
\caption{Layout of the KM2A full array. The red squares and blue dots indicate the EDs and MDs in operation, respectively. Showers are selected with reconstructed cores falling in the ring between the two black circles, featuring inner and outer radii of 320 and 420 m, respectively.}\label{array}
\end{figure}

LHAASO is a ground-based air shower observatory located 4410 m above sea level in Daocheng, China \cite{lhaaodesign}. It contains a hybrid detector array comprising an EAS array covering an area of 1.3 km$^{2}$ (KM2A), 78,000 m$^{2}$ water Cherenkov detector array (WCDA), and 18 wide-field air Cherenkov/fluorescence telescopes (WFCTA). KM2A consists of 1188 MDs with a spacing of 30 m and 5216 electromagnetic detectors (EDs) with a spacing of 15 m; these detectors record muons and electromagnetic particles, respectively, as shown in Fig.~\ref{array}.

Each ED is composed of a plastic scintillation detector with an area of 1 m$^{2}$ covered with a 5 mm lead plate on its surface. It absorbs low-energy charged particles from atmospheric showers and converts gamma rays into positron-electron pairs. EDs are used to detect charged particles and gamma rays in the EAS and to reconstruct the arrival directions and core positions of primary cosmic-rays.

\begin{figure}[htpb]
\centering
\includegraphics[width=7cm]{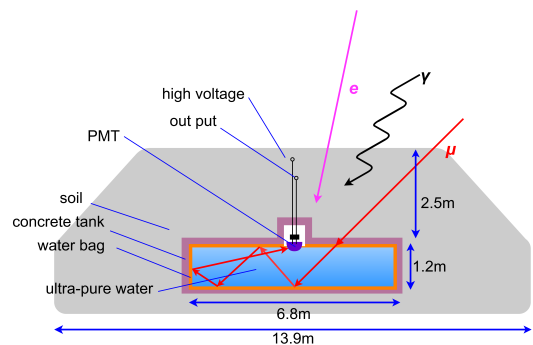}
\caption{Schematic of LHAASO-KM2A muon detector. One water tank is buried under 2.5 m of soil. The ultra-pure water within the tank reaches a depth of 1.2 m and is enclosed in a highly reflective bag.}\label{fig}
\end{figure}

Each MD is composed of a water Cherenkov detector with an area of 36 m$^{2}$. Fig.~\ref{fig} shows a schematic of an MD. Each MD unit consists of a concrete tank with a diameter of 6.8 m and a depth of 1.2 m. The interior of the tank contains ultra-pure water enclosed in a highly reflective ($\geq$95\%) bag. Additionally, there is an overburden soil layer above the MD; it is approximately 2.5 m thick and absorbs electromagnetic and other charged particles in the shower. However, muons with energies greater than 1 GeV can penetrate overburden soil~\cite{PRA2015_KM2Adesignzx}. Thus, an 8-inch photo-multiplier tube (PMT) is positioned at the top surface of the water~\cite{2022yupmt}. It is responsible for collecting the Cherenkov light emitted by high-energy muons. The total muon-sensitive area in LHAASO exceeds 40,000 m$^{2}$. Moreover, owing to its wide dynamic range, the MDs can detect up to 10,000 particles, which enables accurate measurement of muon content across a broad energy range without saturation~\cite{lhaaodesign}.

\begin{figure*}[htpb]
\centering
\begin{minipage}[t]{0.49\textwidth}
\centering
\includegraphics[width=7cm]{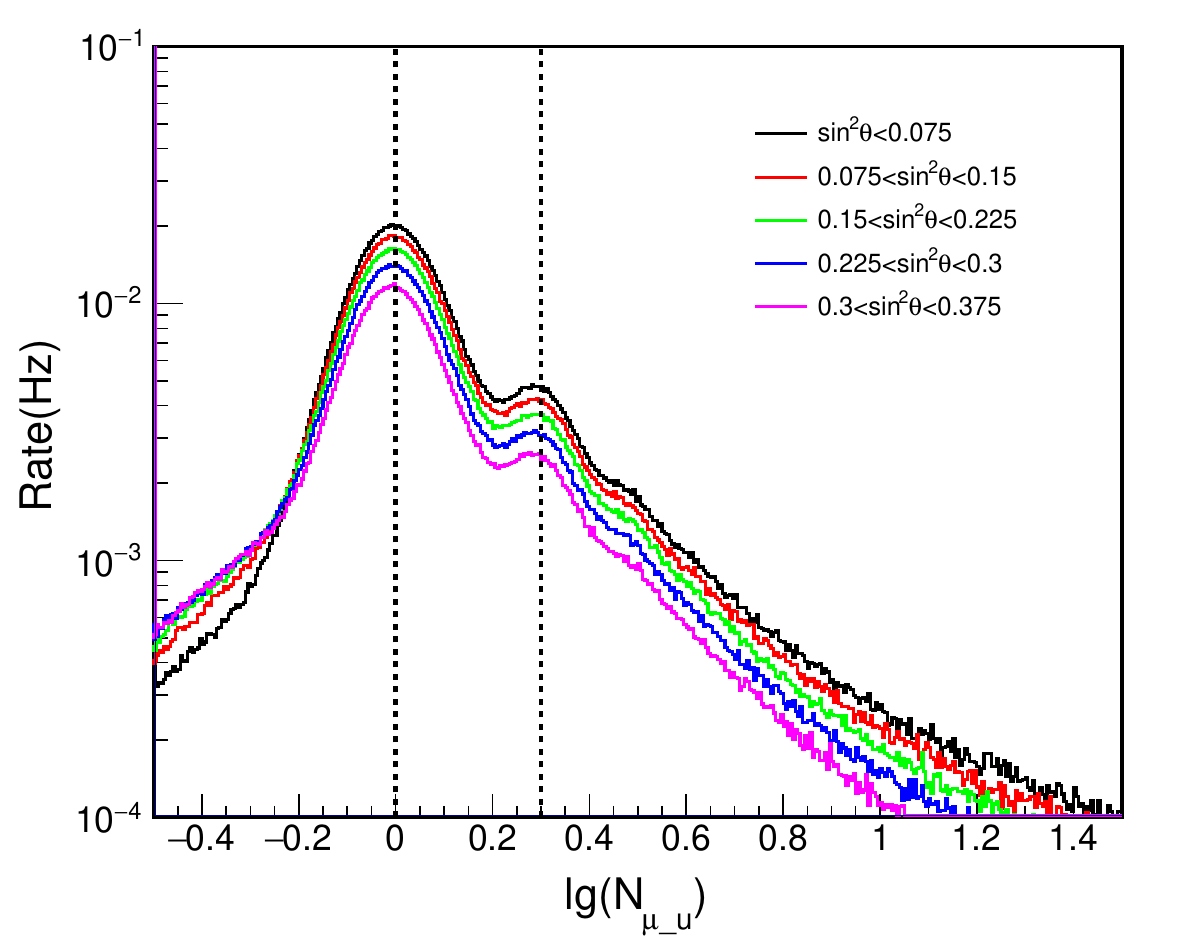}
\end{minipage}
\begin{minipage}[t]{0.49\textwidth}
\centering
\includegraphics[width=7cm]{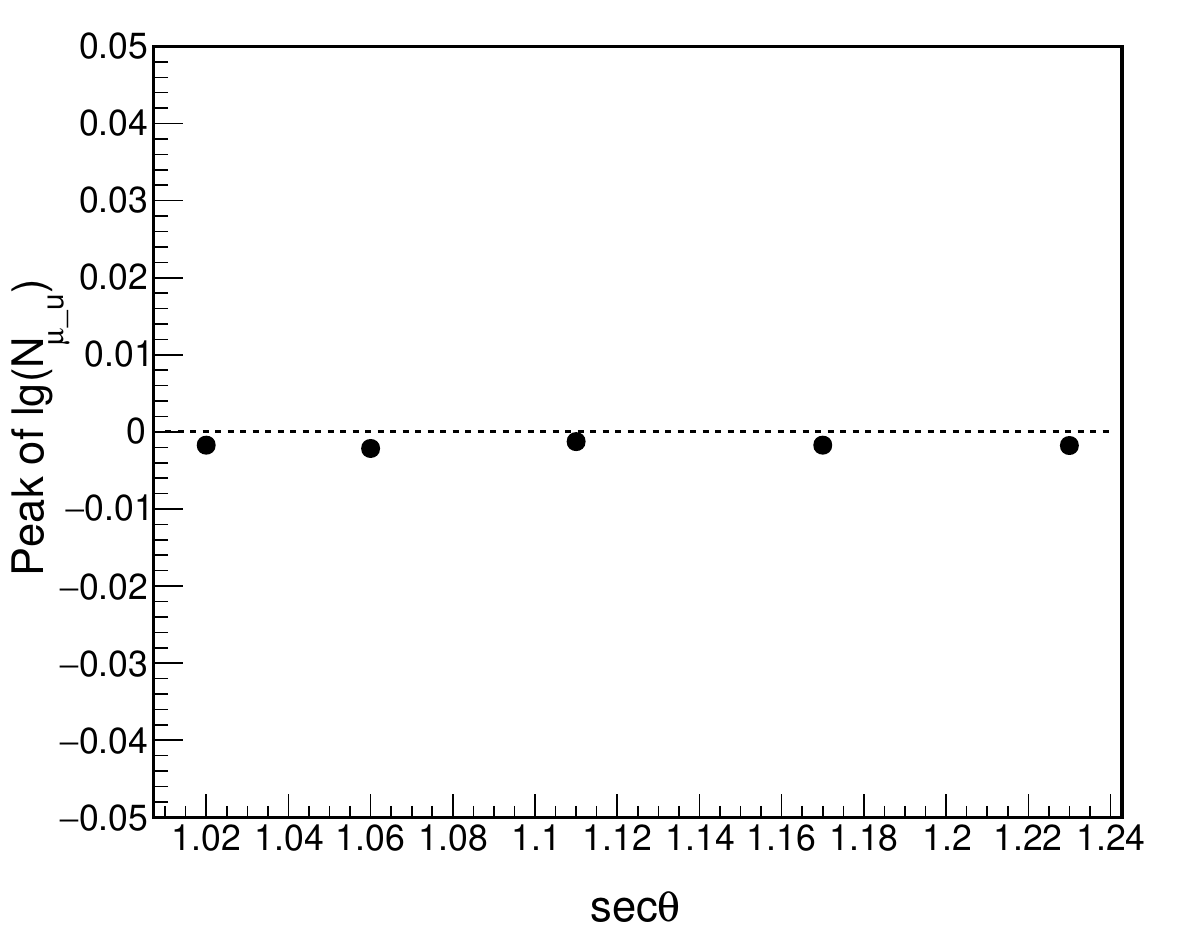}
\end{minipage}
\caption{Left: Logarithmic number of muons distribution of one single MD unit ($N_{\mu\_u}$) after zenith angle correction for each zenith angle. The vertical dash line indicates the position of the single-muon peak and double-muon peak. Right: Fitting value of the logarithmic single-muon peak for each zenith angle; the error bar is smaller than the mark size.}\label{siglenu}
\end{figure*}

\subsection{Muon content measurement\label{sec:Nu}}
When one muon particle passes through an MD, the Cherenkov light emitted from pure water is collected by the PMT after multiple reflections within the tank. The PMT signals pass through a shaping circuit and are then continuously digitized using flash analog-to-digital converters (FADCs)~\cite{lhaaodesign}. When the signal amplitude of one FADC channel exceeds the preset threshold, the signal is recorded as one `hit' and this MD is recorded as a fired detector. An event trigger is generated if more than 20 EDs are fired within a 400 ns time window. In this work, we restrict the arrival time of particles to within [-30, 50] ns around the shower front, to suppress background or detector noise as doing in~\cite{cpclhaasocrab}.

The total number of muons of a triggered event is the sum number of muons from all fired MDs. The number of muons of each fired MD are derived by dividing the number of photonelectrons (NPE) recorded by the MD by VEM/cos($\theta$). Here VEM is the vertical equivalent muon and $\theta$ is the zenith angle of the event. The VEM of each MD is calibrated once daily by almost vertical events~\cite{PRD2022EDcalLv}. An inclined muon travels a longer path in the tank water of the MD compared to a vertical muon, leading to a larger NPE. The NPE produced by a single muon is proportional to the path length; therefore, the equivalent muon for inclined showers should be VEM/cos($\theta$)~\cite{PRA2015_KM2Adesignzx}. The distribution of muons from an individual MD in different directions is shown in the left-hand panel of Fig.~\ref{siglenu}. It can be observed that the single-muon and double-muon peaks are the same for each zenith angle bin. The right-hand panel of Fig.~\ref{siglenu} shows the peak values of single-muon at each zenith angle range more clearly, and no zenith angle dependence was observed.

The high-energy electromagnetic particles near the core of the shower can punch through the shielding soil of an MD and become recorded as muon. To reduce the punch-through effect of high-energy electromagnetic particles near the shower core, hits within 40 m from the shower axis are not counted. Hits 200 m away from the shower axis are not counted either because of the limited size of KM2A. Therefore, in this study, the number of muons recorded by MDs within a distance of 40-200 m from the shower axis in a shower event was defined as the muon content $N_{\mu}$ of a shower event. According to~\cite{PRD2022zhyenergyreconstruction}, the muon content within this 40–200 m range exhibits a strong linear relationship with the true muon content.

\subsection{Monte Carlo simulation}\label{sec2c}

Monte Carlo simulations were performed to compare the results with those obtained using interaction models. The COsmic Ray SImulations for KAscade (CORSIKA) code (version7.7410) \cite{corsikasim} software package was used to simulate EASs of cosmic-rays with primary energies ranging from 100 TeV to 100 PeV. The zenith angles of the incident cosmic-rays were sampled within $0^{\circ}$–$40^{\circ}$. In CORSIKA simulation, the low-energy hadronic interaction model Fluka~\cite{BATTISTONI201510} and high-energy hadronic interaction models QGSJET-II-04~\cite{PhysRevD.83.014018}, EPOS-LHC~\cite{PhysRevC.92.034906}, and SIBYLL 2.3d~\cite{PhysRevD.102.063002} were employed for model validation. The number of events simulated by the QGSJET-II-04 model exceeded 7.1$\times$10$^{7}$, whereas the number of events simulated by EPOS-LHC and SIBYLL 2.3d exceeded 3.5$\times$10$^{7}$. The primary particle components in the simulation included hydrogen (H), helium (He), nitrogen (N), aluminum (Al), and iron (Fe) nuclei.

The ED and MD detector responses were simulated using the G4KM2A \cite{km2asimulationnew} package developed in the framework of GEANT4 \cite{g4toolkit}. Random noise in a single ED and MD was also considered in the simulation.

The simulation data were normalized according to several popular assumed energy spectra, including Gaisser H3a~\cite{Gaisser2013}, Horandel~\cite{HORANDEL2003193}, and Global Spline Fit (GSF)~\cite{ICRCGSF2017}. A new spectrum model~\cite{lv2024precise} developed according to the last published LHAASO spectrum results~\cite{PhysRevLett.132.131002} was used; in this study, it is denoted as the LHAASO spectrum. The simulation results were validated using LHAASO KM2A data, which matched well with the data~\cite{cpclhaasocrab}.

\begin{figure}[htpb]
\centering
\includegraphics[width=7cm]{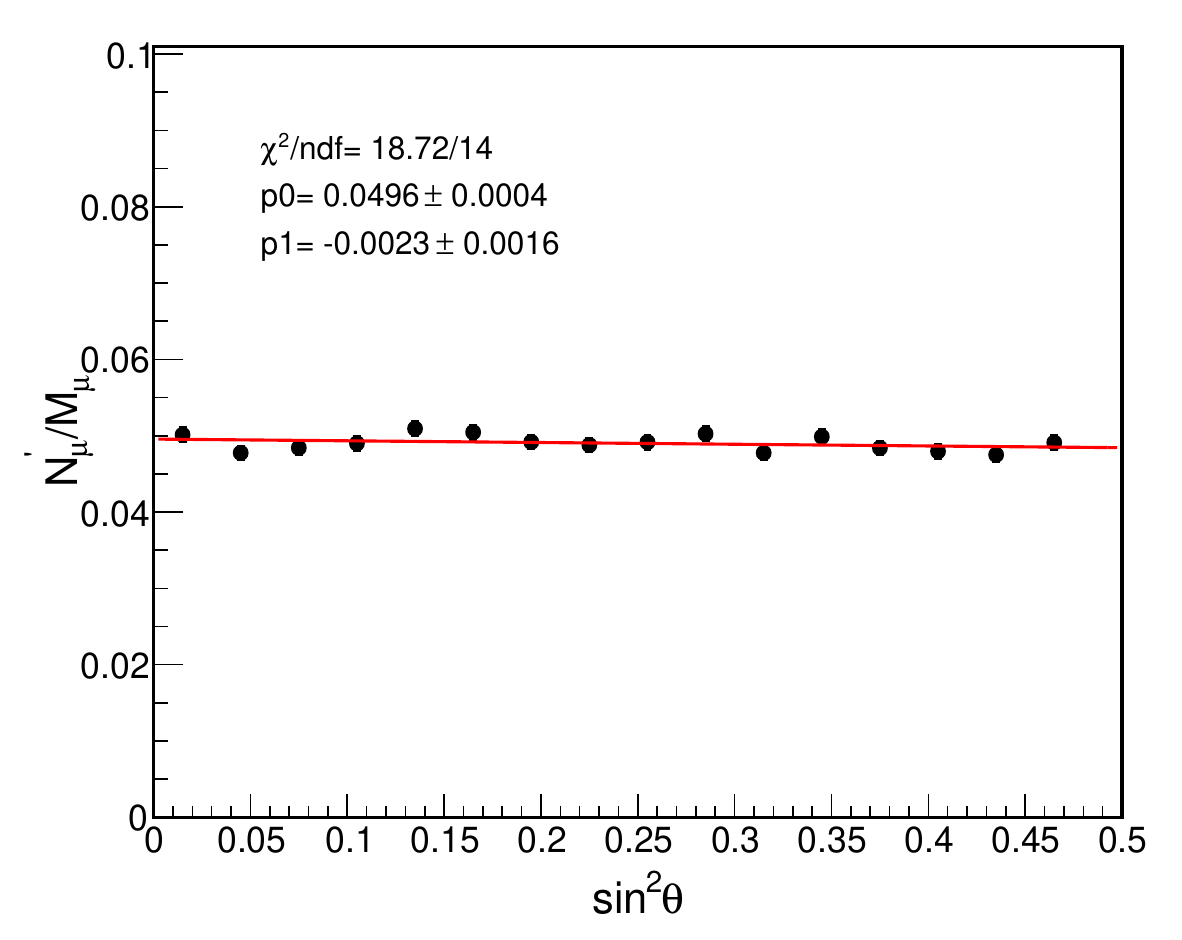}
\caption{The distribution of the ratio between injected and measured number of muons across different zenith angles. The red line represents the result of linear fitting.}\label{nuratio}
\end{figure}

The G4KM2A package, operating in single unit mode, was also used to investigate the detector's response to muon counting across different zenith angle windows. A total of 1.2 million muons were injected within a 15 m radius, with zenith angle uniformly distributed according to sin$^{2}\theta$ in the range of 0$^{\circ}$-45$^{\circ}$. The number of injected-muons within one zenith angle bin is denoted as $M_{\mu}$, and the number of measured-muons within this zenith angle bin is $N_{\mu}^{'}$ (also corrected with VEM/cos($\theta$) as detector done). Fig.~\ref{nuratio} illustrates the ratio $N_{\mu}^{'}/M_{\mu}$ for different zenith angles. The fitting results indicate the ratio remains almost constant across the zenith angle range, indicating that the detector's response to the number of muons is independent of the zenith angle of the air shower.

\subsection{Data and event selection\label{sec:Filter}}

The experimental data spanned two years, from August 2021 to July 2023, totaling 730 days. To obtain high-quality measurement results, the following event selection criteria were applied: 
\begin{enumerate}[label=(\arabic*)]
\item The electromagnetic particle number ($N_{e}$) of the shower, considering the ED within a distance of 40-200 m from the shower axis, must be greater than 80;
\item The distance from the shower core to the array center is within 320-420 m, as shown by the black circles in Fig.~\ref{array};
\item Zenith angle $\theta$ $\leq$ 40$^{\circ}$.
\end{enumerate}

\begin{figure}[htpb]
\centering
\includegraphics[width=7cm]{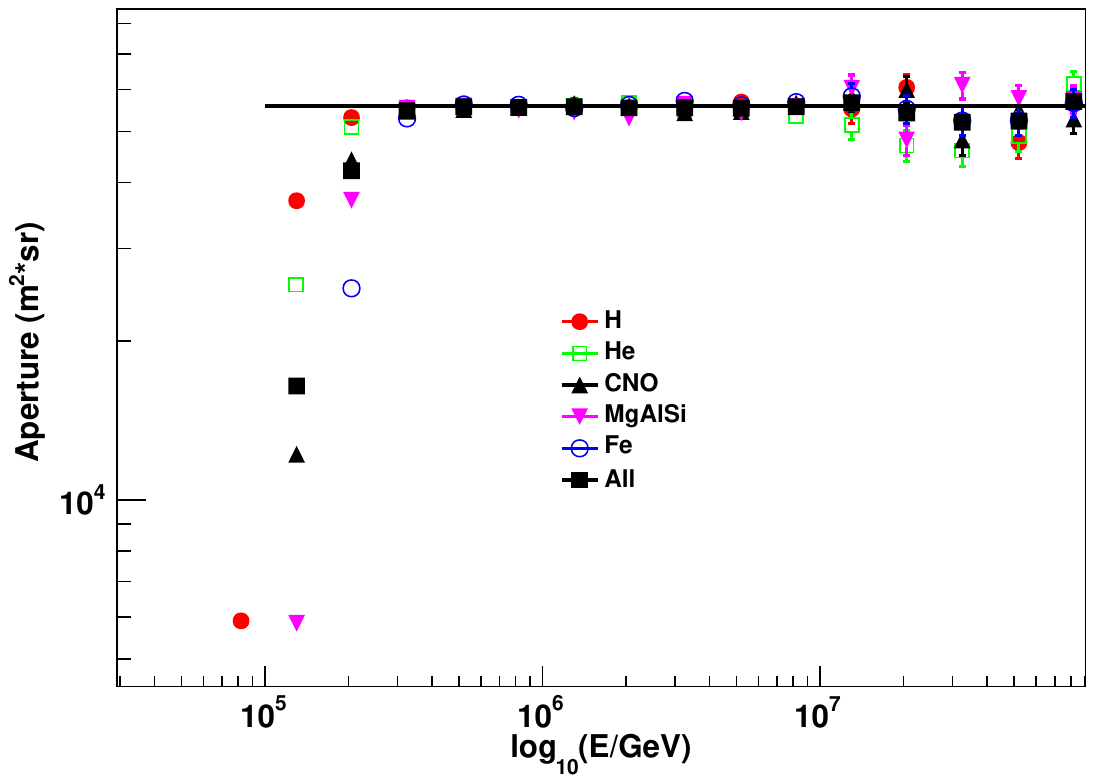}
\caption{The aperture of the KM2A full-array for cosmic-rays varies with the primary energy of different components. The selection criteria were $N_{e}$ $\geq$ 80, shower core within 320-420 m, and 33.41$^\circ$ $\leq$ $\theta$ $\leq$ 38$^{\circ}$.}\label{aperture}
\end{figure}

\begin{figure*}[htpb]
\centering
\begin{minipage}[t]{0.49\textwidth}
\centering
\includegraphics[width=7cm]{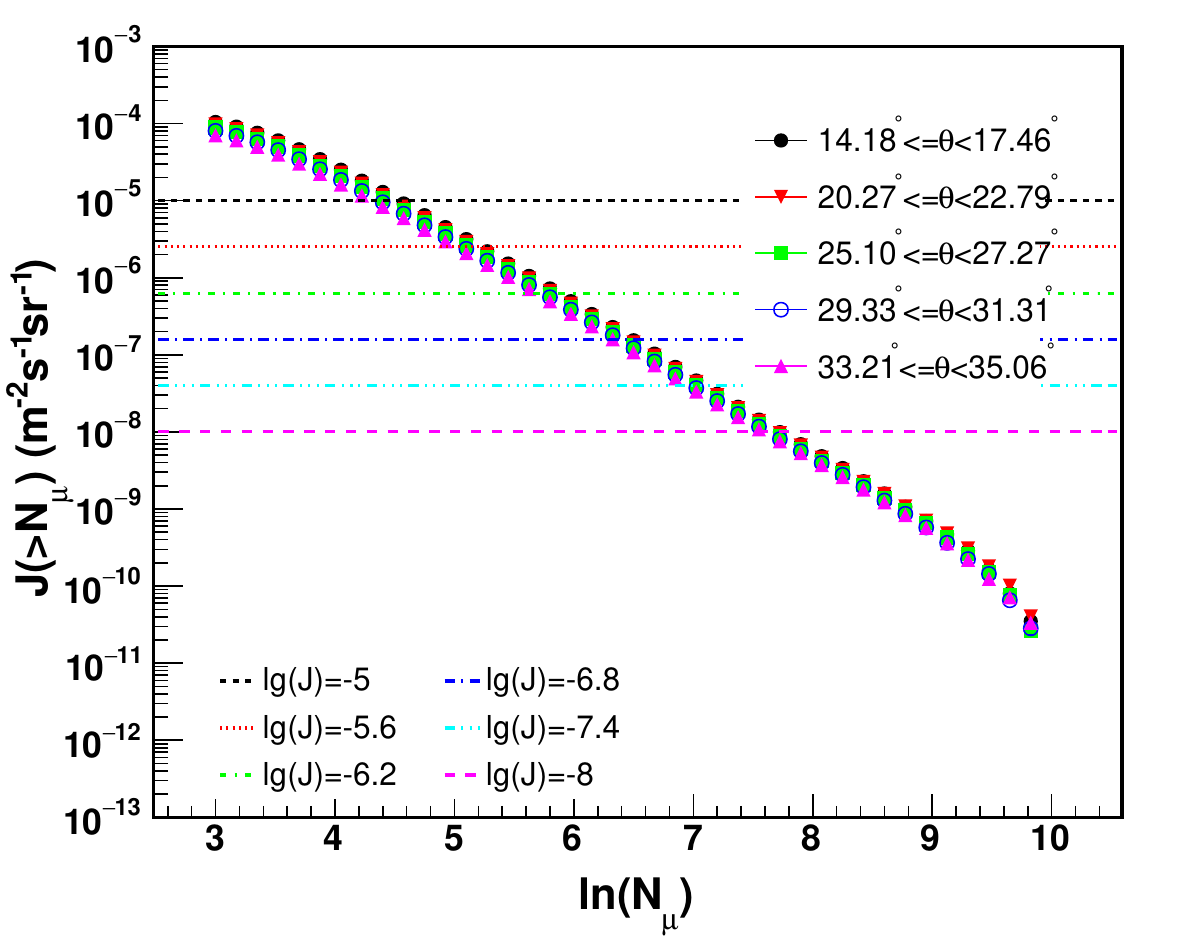}
\end{minipage}
\begin{minipage}[t]{0.49\textwidth}
\centering
\includegraphics[width=7cm]{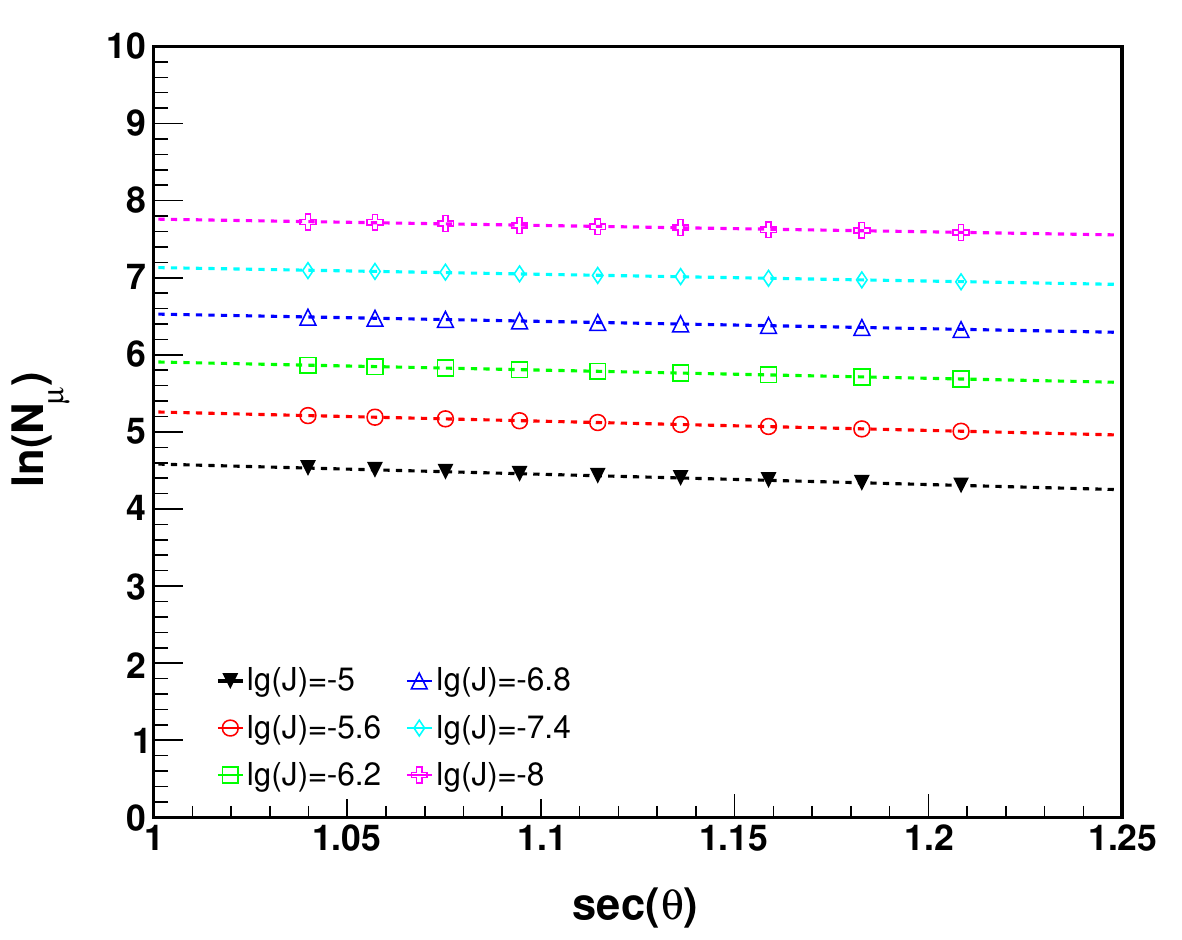}
\end{minipage}
\caption{Left: Distribution of the integral muon flux for five zenith angle intervals with same aperture. The different marks represent the integral spectra for each zenith angle intervals. The dashed lines of six different colors represent the six cut fluxes. The statistical error bar is shorter than the mark size. Right: Logarithmic muon content distribution in terms of the zenith angle of the shower corresponding to the CIC method at one constant flux. The six different colored marks represent the variations of $N_{\mu}$ with zenith angle under six different cut fluxes. The error bars are smaller than the size of the marks.}\label{att}
\end{figure*}

The effective detection area of KM2A can be calculated using simulation data and is dependent on the energy, zenith angle, and composition. Fig.~\ref{aperture} shows the apertures of different components within the largest zenith angle bin in this study (between 33.41$^\circ$ and 38$^\circ$) after event selection. The effective aperture increases with energy and reaches full efficiency for all components when the shower energy exceeds lg($E$/GeV)$\geq$5.5. This energy threshold also guarantees full efficiency for showers with zenith angles smaller than those shown in this plot, given that the efficiency increases with decreasing zenith angle. The subsequent analysis is based on full-efficiency measurement and requires a primary energy lg($E$/GeV)$\geq$5.5.

The atmospheric pressure has an impact on the measurement of muon content. By utilizing the relationship between the number of muons and the atmospheric pressure, we adjusted the number of muons to that under standard atmospheric pressure. The influence of atmospheric pressure on the measurement of muon content was 4.4$\%$, becoming 0.6$\%$ after correcting for atmospheric pressure~\cite{PhysRevLett.132.131002}.

The calibration uncertainty of one ED and MD for a single particle are less than 2$\%$~\cite{PRD2022EDcalLv} and 0.5$\%$~\cite{PRA2015_KM2Adesignzx}, respectively. At 300 TeV, the average numbers of triggered EDs and MDs were approximately 200 and 50, respectively. Therefore, the measurement uncertainties of $N_{e}$ and $N_{\mu}$ are both less than 0.15$\%$. The zenith angle resolution is less than 0.2$^{\circ}$ at 400 TeV~\cite{cpclhaasocrab}.

\section{Measurement of the attenuation length of muon content\label{sec:Three}}

\subsection{Attenuation length measurement}\label{sec:31}

The CIC method~\cite{CICmothed} was used to infer the attenuation length of muon content in air showers. This method is based on the assumption that the arrival distribution of cosmic-rays is isotropic; thus, the observed intensity of primary particles with the same energy is independent of the zenith angle or slant depth. This method depends on no supposed simulation models. The purpose of this study was to compare the muon content in air showers with the same primary energy at different zenith angles or different atmospheric depths.

To apply the CIC method, the data after selection were grouped into nine zenith angle intervals (zenith angle within 14.18$^\circ$-35.06$^\circ$ with the same aperture $\Delta$sin$^{2}\theta$=0.03, shower core within 320-420 m) as shown in the left-hand panel of Fig.~\ref{att}, where only five zenith angles are displayed. The integral muon flux, $J(\geq N_{\mu},\theta)$ (m$^{-2}$$\cdot$s$^{-1}$$\cdot$sr$^{-1}$), was estimated based on the differential muon content spectrum using the following formula:

\begin{equation}
J(\geq N_{\mu},\theta)= \int_{N_{\mu}}{}\Phi(N_{\mu},\theta) dN_{\mu} ,
\label{eq2}
\end{equation}

\noindent
where $\Phi$(N$_{\mu}$,$\theta$) represents the differential muon content spectrum in the direction of the zenith angle $\theta$. CICs were performed at a constant flux $J(\geq N_{\mu},\theta)$ on the integral spectrum of each zenith angle $\theta$ direction within fourteen flux ranges: lg[$J$/(m$^{-2}$$\cdot$s$^{-1}$$\cdot$sr$^{-1}$)]$\in$[-8.9, -5.0]. The logarithmic muon contents of the CICs in the flux along direction $\theta$ are shown in the right-hand panel of Fig.~\ref{att} for each flux cut; only six cut values are displayed.

At each flux cut, the muon content of the air shower was attenuated as the zenith angle $\theta$ increased (with a longer atmospheric depth), as shown in the right-hand panel of Fig.~\ref{att}. To obtain the attenuation length of the muon content, we fit the attenuation curves using the following formula:

\begin{equation}
N_\mu(\theta) = N^0_\mu e^{\frac{-X_0 \cdot sec(\theta)}{\Lambda_{\mu}}} ,
\label{eq1}
\end{equation}

\noindent
where $N_{\mu}(\theta)$ is the muon content of the shower with zenith angle $\theta$; $N^{0}_{\mu}$ is the normalization parameter, which is equal to the muon content of the vertical incident-event showers; $X_0$ is the vertical atmospheric depth, equal to 600 g$\cdot$cm$^{-2}$ at LHAASO, and $\Lambda_{\mu}$ is the attenuation length of the muon content.

\begin{figure}[!htpb]
\centering
\includegraphics[width=7cm]{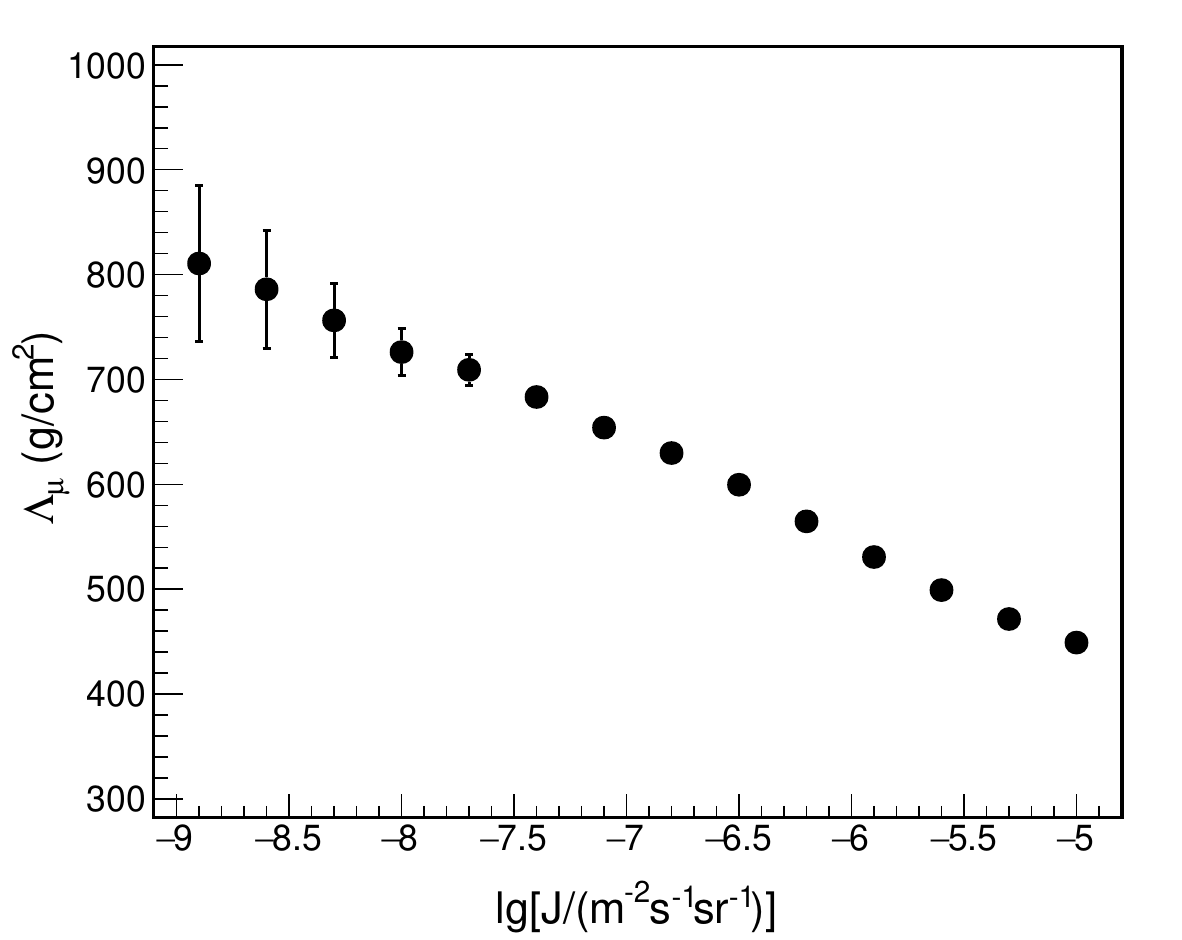}
\caption{The attenuation length of the muon content varies with the flux cut. The bars represent statistical errors.}\label{dataju}
\end{figure}

The attenuation length $\Lambda_{\mu}$ can be obtained by linearly fitting the relationship between the logarithmic muon content (ln($N_{\mu}$)) and zenith angle (sec$(\theta)$) or atmospheric depth (X$_{0}\cdot$sec$(\theta)$). The slope of the fitted line is expressed as $X_{0}/\Lambda_{\mu}$. The attenuation lengths of the muon content for different flux cuts were measured according to the fitting results, as shown in Fig.~\ref{dataju}. The attenuation length of the muon content increased from 449 to 811 g$\cdot$cm$^{-2}$ for the logarithmic integral flux decreasing from -5.0 to -8.9 m$^{-2}$$\cdot$s$^{-1}$$\cdot$sr$^{-1}$. The attenuation length for each flux cut is listed in Table~\ref{table1} together with the corresponding statistical error. These results, which constitute the baseline for the systematic error analysis presented in the next section, are independent of any supposed hadronic interaction or spectrum models.

\subsection{Systematical errors}\label{sec33}

Atmospheric pressure variation results in fluctuations of the atmosphere depth and muon content. However, it becomes negligible after correction of the muon content with the atmospheric pressure, as discussed in part D of Section~\ref{sec:TWO}. The measurement precision of the muon content was very high, as mentioned in that part. Therefore, the uncertainties induced by the atmospheric pressure and muon content precision were disregarded in this study.

The measured results were based on the integral flux of ln($N_{\mu}$), and the effect of the bin size of ln($N_{\mu}$) on the attenuation lengths was investigated. Additionally, the number of degrees of freedom during the fitting process was examined. These systematic errors were too small to be considered.

The main systematic errors were as follows:

\begin{itemize}
    \item[(1)] Zenith angle interval size
    \item[(2)] Zenith angle range
    \item[(3)] Shower core range
\end{itemize}

In the baseline analysis above, events within a zenith angle range of $14.18^{\circ}$–$35.06^{\circ}$ were divided into nine zenith angle intervals based on $\Delta$sin$^{2}\theta$=0.03. For comparison, the zenith angle was divided into intervals with a larger aperture $\Delta$sin$^{2}\theta$=0.05 and smaller aperture $\Delta$sin$^{2}\theta$=0.02. The uncertainties in the measured attenuation length from the zenith angle interval size are listed in the A(zenith angle interval size)-th column of Table~\ref{table1} for each flux cut. The uncertainties are less than 2$\%$ for flux lg[$J$/(m$^{-2}$$\cdot$s$^{-1}$$\cdot$sr$^{-1}$)]$\geq$-8.0, and the uncertainty is larger than 10$\%$ for the last point lg[$J$/(m$^{-2}$$\cdot$s$^{-1}$$\cdot$sr$^{-1}$)]=-8.9, but still comparable with its statistical error.

In the baseline analysis, the range of the fitted zenith angle was $14.18^{\circ}$–$35.06^{\circ}$, corresponding to the atmospheric depth (X$_{0}$$\cdot$sec$(\theta)$) within 618-733 g$\cdot$cm$^{-2}$. We skipped the observation aperture up and down in two zenith angle ranges, namely $0^{\circ}$–$31.31^{\circ}$ and $20.27^{\circ}$–$38.65^{\circ}$ with equal apertures, and measured the attenuation length for the same zenith angle interval as the baseline. The uncertainties from the zenith angle range are listed in the B(zenith angle range)-th column of Table~\ref{table1} for each flux cut. The uncertainties are similar to those caused by the size of the interval and were controlled for the flux lg[$J$/(m$^{-2}$$\cdot$s$^{-1}$$\cdot$sr$^{-1}$)]$\geq$-8.3.

In the baseline analysis, the shower core of the events was limited to 320-420 m from the center of the array. A wider (narrower) shower core range of 300-450 (350-400) m from the array center was selected for systematic error study. The uncertainties in the shower core selection are listed in the C(shower core range)-th column of Table~\ref{table1} for each flux cut. The uncertainties were almost the same for all fluxes except for the last one. According to Table~\ref{table1}, the shower core is the dominant uncertainty for a larger flux but is equal to the above two systematic uncertainties when the flux lg[$J$/(m$^{-2}$$\cdot$s$^{-1}$$\cdot$sr$^{-1}$)]$\leq$-8.6. 

\begin{figure}[!htpb]
\begin{minipage}[t]{0.48\textwidth}
\centering
\includegraphics[width=7cm]{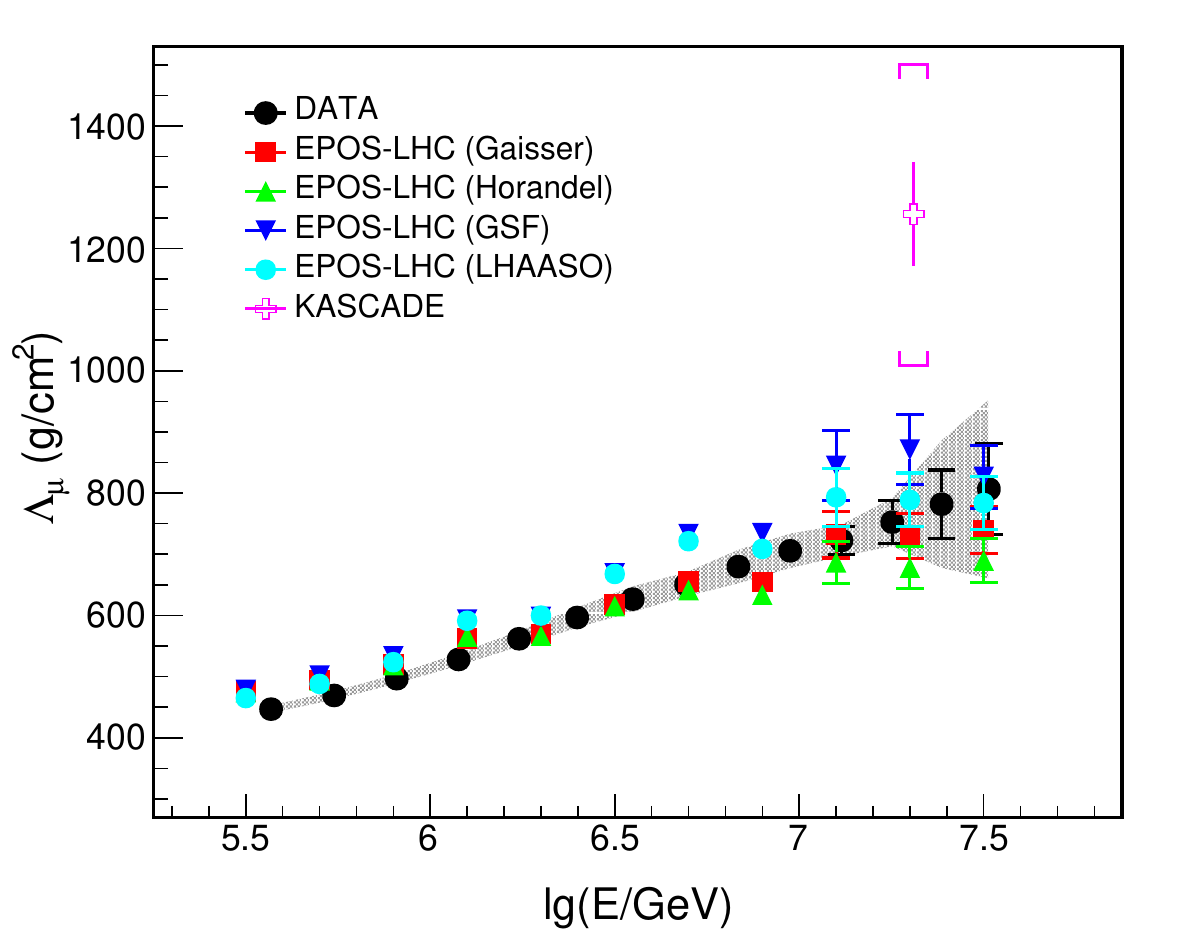}\\
\end{minipage}
\begin{minipage}[t]{0.48\textwidth}
\centering
\includegraphics[width=7cm]{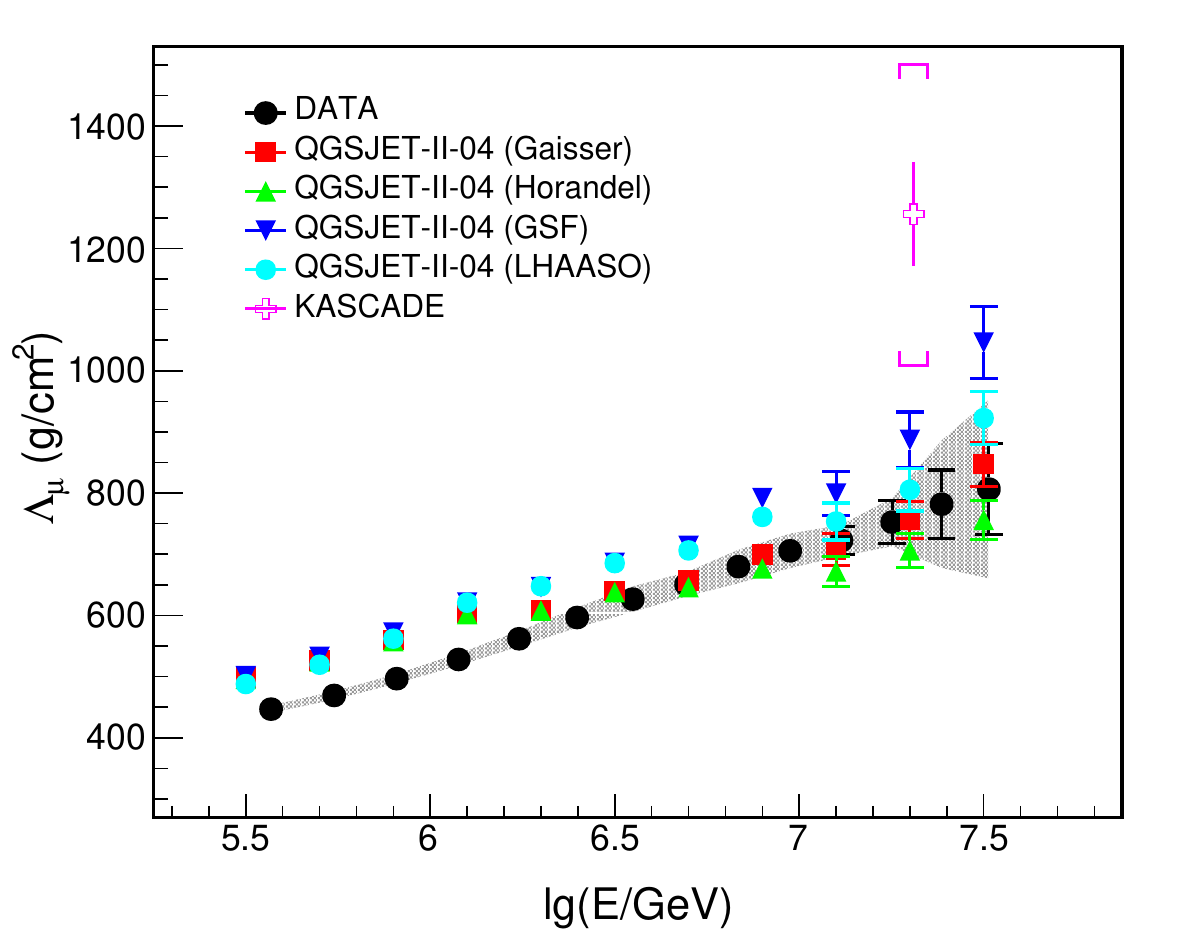}\\
\end{minipage}
\begin{minipage}[t]{0.48\textwidth}
\centering
\includegraphics[width=7cm]{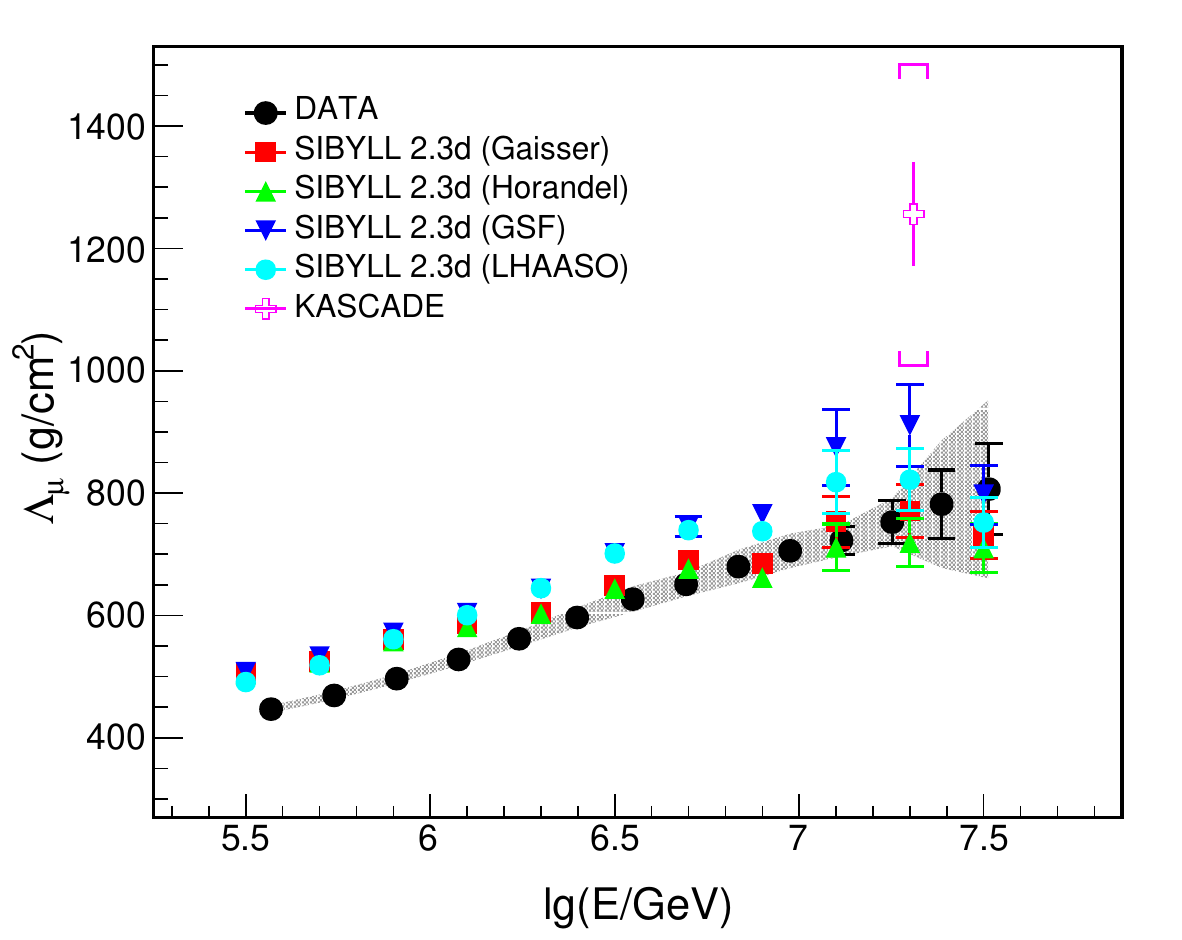}\\
\end{minipage}
\caption{The muon content attenuation length varies with the cosmic-ray energy. The measured results with LHAASO data are represented by black dots. The error bars represent the statistical error whereas the gray shadow represents the systematic errors. The error bar of the energy is shorter than the dot size. The attenuation lengths are plotted according to the simulation data produced with the hadronic interaction models EPOS-LHC (upper panel), QGSJET-II-04 (middle panel), and SIBYLL 2.3d (down panel), after normalization with spectrum models. Only the statistical errors are shown for the simulation results. The red square, green up triangle, and blue down triangle represent the attenuation length of the muon content for the spectrum models of the Gaisser H3a~\cite{Gaisser2013}, Horandel~\cite{HORANDEL2003193}, and GSF~\cite{ICRCGSF2017} models, and the azure dots represent the results from the LHAASO spectrum model, which was developed based on the all-particle energy spectrum~\cite{lv2024precise} measurement of LHAASO. The pink cross represents the attenuation length of the KASCADE results~\cite{KASCADEatt} with energy ranging from 20 to 100 PeV.}\label{sumatt}
\end{figure}

The last column in Table~\ref{table1} lists the total systematic errors for each flux cut. The total systematic errors were less than 6$\%$ except for the last two flux cuts (lg[$J$/(m$^{-2}$$\cdot$s$^{-1}$$\cdot$sr$^{-1}$)]$\geq$-8.6). The total systematic errors were larger than the statistical errors for all the flux cuts.

\subsection{Comparison of experimental and simulated attenuation lengths}\label{sec32}

The LHAASO Collaboration has recently published the most accurate all-particle differential energy spectrum within the energy range of 0.3-30 PeV~\cite{PhysRevLett.132.131002}. The all-particle energy spectrum of LHAASO fitted well with the following formula:

\begin{equation}
J (E) =\Phi_{0}\cdot(E)^{\gamma_{1}} \left( 1+\left(\frac{E}{E_{b}}\right)^{s} \right) ^{(\gamma_{2}-\gamma_{1})/s} ,
\end{equation}\label{enerspetrum}

\noindent
where $E_{b}$=3.67 PeV corresponds to the knee position, $\gamma_{1}$=-2.7413 and $\gamma_{2}$=-3.128 are the slopes of the energy spectrum below and above the knee, and $s$=4.2 is the sharpness of the knee.

By integrating this spectrum, the corresponding shower energy of each cosmic-ray flux was estimated and listed in the second column of Table~\ref{table1}. The uncertainty of this estimated energy is less than 3$\%$ according to the uncertainty of the published energy spectrum, which includes the systematic error from the interaction models, composition models, and atmospheric pressure.

Therefore, the flux is transferred to cosmic-ray energy according to Table~\ref{table1}, and the variation in the attenuation length with shower energy is shown by the black dots in Fig.~\ref{sumatt}. The attenuation lengths of the muon content increase with the cosmic-ray energy. The systematic error is also shown as a shadow in this figure, according to the total systematic error shown in Table~\ref{table1}.

To validate the interaction models with our measured results, cosmic-ray air showers were simulated using three hadronic interaction models. The simulation data were normalized based on four energy spectrum models, as described in part C of Section~\ref{sec:TWO}. Utilizing these normalized cosmic-ray air shower simulation data, the attenuation lengths of the muon content were re-evaluated by fitting the variation in muon content with the zenith angle for each shower energy bin, similar to the data. The simulation results are plotted together with results from LHAASO data in Fig.~\ref{sumatt} to compare the events with equal energy for three hadronic interaction models: EPOS-LHC, QGSJET-II-04, and SIBYLL 2.3d. The results of four energy spectrum models (Gaisser H3a, Horandel, GSF, and LHAASO) were compared for each interaction model.

The increasing trend in the attenuation lengths of the muon content with the cosmic-ray energy was confirmed by the simulation results for the three hadronic interaction models, as shown at the top, middle, and bottom panels of Fig.~\ref{sumatt}. The attenuation lengths of the simulation data for both QGSJET-II-04 and SIBYLL 2.3d models were significantly longer than those of the LHAASO data, as shown in the middle and bottom panels of Fig.~\ref{sumatt}. The results of the EPOS-LHC model were closer to the experimental data than those of the other models, as shown in the top panel of Fig.~\ref{sumatt}. In future analyses, we will continue to study the variations according to different hadronic models relative to energy.

The attenuation lengths obtained from the LHAASO data were shorter than those obtained from the simulation data, which were normalized to the LHAASO energy spectrum over the entire energy range. The results for the GSF energy spectrum were similar to those of the LHAASO energy spectrum. However, the attenuation lengths of the Gaisser H3a and Horandel energy spectra were shorter than those of the LHAASO and GSF spectra above 2 PeV, where the mean mass of the Gaisser H3a and Horandel models is heavier than that of the data~\cite{PhysRevLett.132.131002}.

The KASCADE-Grande Collaboration measured the attenuation length of the muon content for cosmic-ray air showers within 20-100 PeV at sea level. In particular, the muon content was measured by fitting the lateral distribution with a Lagutin-Raikin function with fixed shape. The attenuation length was 1256$\pm$85$_{-232}^{+229}$ g$\cdot$cm$^{-2}$, which is longer than the results from simulation data~\cite{KASCADEatt}. The results from LHAASO data with similar energy were shorter than that from KASCADE data, as shown in Fig.~\ref{sumatt}, but close to the simulation results. 

\begin{table*}[htpb]
\centering
\caption{List of attenuation lengths of the muon content for each flux (shower energy) with systematical and statistical errors; the A-th column represents the zenith angle interval size, the B-th column represents the zenith angle range, and the C-th column represents the shower core range.}
\begin{tabular}{*{8}{cclccccc}}
\hline\hline
 lg$[J/$(m$^{-2}$$\cdot$s$^{-1}$$\cdot$sr$^{-1}$)] & lg[$E_{rec}$/(GeV)] & $\Lambda_{\mu}$ $\pm$ Stat $\pm$ Syst (g$\cdot$cm$^{-2}$)& A($\%$) & B($\%$) & C($\%$) & Stat($\%$) & Syst($\%$) \\
\hline
-5.0 & 5.57 & 449.1$\pm$0.3$\pm$6.7 & 0.1 & 0.6 & 1.3 & 0.1 & 1.5 \\
-5.3 & 5.74 & 471.6$\pm$0.4$\pm$7.1 & 0.2 & 0.4 & 1.4 & 0.1 & 1.5\\
-5.6 & 5.91 & 499.2$\pm$0.7$\pm$8.0 & 0.4 & 0.4 & 1.5 & 0.2 &1.6\\
-5.9 & 6.08 & 530.7$\pm$1.1$\pm$10.1 & 0.6 & 0.8 & 1.6 & 0.2 & 1.9\\
-6.2 & 6.24 & 564.7$\pm$1.7$\pm$15.8 & 0.8 & 1.3 & 1.8 & 0.3 & 2.3\\
-6.5 & 6.40 & 599.6$\pm$2.6$\pm$16.8 & 1.1 & 1.8 & 1.8 & 0.5 &2.8\\
-6.8 & 6.55 & 629.9$\pm$4.0$\pm$21.4 & 1.9 & 2.2 & 1.8 & 0.7 & 3.4\\
-7.1 & 6.69 & 654.1$\pm$6.0$\pm$20.3 & 1.1 & 2.3 & 1.8 & 1.0 &3.1\\
-7.4 & 6.84 & 683.3$\pm$9.2$\pm$28.0 & 1.9 & 2.6 & 2.6 & 1.4 & 4.1\\
-7.7 & 6.98 & 709.2$\pm$14.6$\pm$29.1 & 1.7 & 2.1 & 3.0 & 2.1 & 4.1\\
-8.0 & 7.11 & 726.1$\pm$22.5$\pm$26.1 &1.7 & 2.1 & 2.4 & 3.1 & 3.6\\
-8.3 & 7.25 & 756.3$\pm$35.2$\pm$40.1 & 3.6 & 3.3 & 2.1 & 4.7 & 5.3\\
-8.6 & 7.39 & 786.0$\pm$56.2$\pm$110.0 & 7.0 & 11 & 3.4 & 7.2 & 14\\
-8.9 & 7.51 & 810.6$\pm$74.6$\pm$153.9 & 12 & 12 & 7.5 & 9.2 & 19\\

\hline\hline
\end{tabular}
    \label{table1}
\end{table*}

\section{Conclusion\label{sec:FOUR}}

LHAASO KM2A has measured electromagnetic particles and muons with very high precision for cosmic-ray air showers with energy between 0.3-30 PeV, where the cosmic-ray energy spectrum becomes steeper. By using the data collected by LHAASO-KM2A within two years, from August 2021 to July 2023, the attenuation lengths of the muon content for cosmic-ray air showers with energy between 0.3 to 30 PeV were measured using the CIC method. The attenuation length increased from 449 to 811 g$\cdot$cm$^{-2}$ as the shower energy increased from 0.3 to 30 PeV. These results are independent of the proposed models.

The air shower simulations were performed using three hadronic interaction models, namely EPOS-LHC, QGSJET-II-04, and SIBYLL 2.3d, and were normalized according to four energy spectra, including one LHAASO energy spectrum. The simulation results confirmed an increasing trend in the attenuation length of the muon content with the shower energy. The attenuation lengths of the QGSJET-II-04 and SIBYLL 2.3d models were longer than those of the LHAASO data, and the results of EPOS-LHC were relatively close to the LHAASO data. The LHAASO results are more suitable for the EPOS-LHC model than for the other two models.

\section{Acknowledgments}

We would like to thank all staff members who work at the LHAASO site 4400 m above sea level year-round to maintain the detector and water recycling system, electricity power supply, and other components of the experiment operating smoothly. We are grateful to the Chengdu Management Committee of Tianfu New Area for their constant financial support for research using the LHAASO data. We appreciate the computing and data service support provided by the National High Energy Physics Data Center for the data analysis. This research study was supported by the following grants: National Natural Science Foundation of China (Nos.12175121, 12275280, 12393851, 12393852, 12393853, 12393854, 12205314, 12105301, 12305120, 12261160362, 12105294,U1931201, and 12375107, Thailand by the National Science and Technology Development Agency (NSTDA), and the National Research Council of Thailand (NRCT) under the High-Potential Research Team Grant Program (N42A650868).

\section{Author contributions}

X.T. Feng led the drafting of the manuscript and analyzed the data. H.Y. Zhang played a pivotal role in crosschecking, preparing the simulation data, and revising the manuscript. L.L. Ma and C.F. Feng, as co-proponents, provided guidance for data analysis and manuscript revision. Z. Cao is the spokesperson for the LHAASO Collaboration and principal investigator of the LHAASO project. All other authors participated in data analysis, including detector calibration, data processing, event reconstruction, data quality check, and various simulations, and provided comments on the manuscript.


\bibliography{reference.bib}
\nocite{}

\end{document}